\documentclass{LMCS}

\usepackage{venanzio,amsfonts,amsmath,amssymb}
\usepackage{hyperref}

\def\doi{1 (2:1) 2005}
\lmcsheading%
{\doi}
{28}
{}
{}
{Dec.~22, 2004}
{Jul.~13, 2005}
{}
\begin{document}

\title{General Recursion via Coinductive Types}
\author[V.\ Capretta]{Venanzio Capretta}
\address{
Department of Mathematics and Statistics\\
University of Ottawa\\
585 King Edward Avenue, Room~204\\
Ottawa, ON, Canada, K1N 4N5}
\email{venanzio.capretta@mathstat.uottawa.ca}


\bibliographystyle{plain}

\begin{abstract}
A fertile field of research in theoretical computer science investigates the representation of general recursive functions in intensional type theories.
Among the most successful approaches are: the use of wellfounded relations, implementation of operational semantics, formalization of domain theory, and inductive definition of domain predicates. 
Here, a different solution is proposed: exploiting coinductive types to model infinite computations.
To every type $A$ we associate a type of partial elements $\copar{A}$, coinductively generated by two constructors: the first, $\return{a}$ just returns an element $a\colon A$; the second, $\step{x}$, adds a computation step to a recursive element $x\colon \copar{A}.$
We show how this simple device is sufficient to formalize all recursive functions between two given types.
It allows the definition of fixed points of finitary, that is, continuous, operators.
We will compare this approach to different ones from the literature.
Finally, we mention that the formalization, with appropriate structural maps, defines a strong monad.
\end{abstract}

\maketitle

\section{Introduction}
Type theory is---we often claim---a rich functional programming language with dependent types and, at the same time, a constructive logical system.
This view is consistently adopted in Martin-L\"of's type theory \cite{martin-loef:1982,nps:1990}.
However, a serious limitation of type theory with respect to standard functional programming languages is that all computations must terminate.
This restriction has two reasons.
First, to decide type-checking of dependent types, we need to reduce type expressions to normal form (see the work by Barendregt and Geuvers \cite{barendregt/geuvers:2001} for a good exposition of technical issues of type-theoretic proof assistants).
Second, since propositions are represented by types and proofs by programs, according to the {\em Curry-Howard isomorphism\/} \cite{howard:1980,geuvers:1993,barendregt:1992,sorensen/urzyczyn:1998}, we cannot allow non-terminating proofs, because they would lead to inconsistency.

There have been attempts to overcome both problems in the literature, starting from the work of Paulson \cite{paulson:1986b} and  Nordstr\"om \cite{nordstrom:1988}.

There are implementation of {\em extensional\/} type theory for which the type-checking problem is undecidable, notably Nuprl \cite{constable:1986}.
In such a system the correctness of a judgment cannot be determined by just analyzing the terms, but a supplementary proof must be given.
In principle, a complete type-checkable term can be generated from such proof.
Types of partial functions have been devised by Constable and added to Nuprl \cite{constable:1983,constable/smith:1987}.

Geuvers, Poll, and Zwanenburg \cite{gpz:1999} added a fixed point operator Y to type theory and proved the conservativity of the extension.
This means that a judgment derived in the extended system and not containing Y can be derived in the non-extended system.
This allows the use of Y for proof search: We construct a proof containing Y, then we reduce it; if we obtain a term not containing Y, we have a valid proof.
However, statements about diverging elements cannot be proved, because valid proofs cannot contain Y.

Recent research has tried to find good representations of general recursive functions in type theory following various avenues:
Balaa and Bertot used well-founded recursion \cite{balaa/bertot:2000};
Dubois and Donzeau-Gouge and, independently, Bove and Capretta used inductive characterizations of domain predicates \cite{dubois/donzeau:1998,bove:2001,bove/capretta:2001,bove/capretta:2004a,bove/capretta:2005a};
Bertot, Capretta, and Das Barman combined the two methods to give a semantics of imperative programming \cite{bcd:2002} and Bertot extended the work to coinductive types \cite{bertot:2005};
Barthe and others used type labeling to strengthen termination conditions \cite{BFGPU:2002,bgp:2005};
McBride and McKinna used {\em views}, that is, different potential inductive characterizations of data types \cite{mcbride/mckinna:2002}.

The first problem is how to represent partial functions.
The most popular way consists in seeing a partial function $f\colon A\pararrow B$ as a total function defined on a subset of the type $A$.
In intensional type theory, a subset of a type is represented by a predicate $D\colon A \rightarrow \prop$.
A partial function is represented by a term $f\colon \Pi x\colon A.(D\ x)\rightarrow B$, provided that the result is not dependent on the proof of the predicate, that is, for $x\colon A$, if $p_1$ and $p_2$ are two proofs of $(D\ x)$, then $(f\ x\ p_1)=(f\ x\ p_2)$.

This formalization of partial functions has two defects.
First, if we want to define the type of partial functions from $A$ to $B$, we need second order sum types:
$$
A\pararrow B \equiv
\begin{array}[t]{l}
\Sigma D\colon A\rightarrow \prop.\\
\Sigma f\colon \Pi x\colon A.(D\ x)\rightarrow B.\\
\forall x\colon A.\forall p_1,p_2\colon (D\ x).
         (f\ x\ p_1)=(f\ x\ p_2).
\end{array}
$$
This requires either an impredicative type theory \cite{girard:1972,girard:1989,coquand/huet:1988} or a predicative theory with type universes \cite{martin-loef:1984}.
In the second case, there are some difficulties when we try to define the predicate inductively from the recursive equations of the function: The counterexample $\mathsf{Itz}$ in \cite{bove/capretta:2005a} cannot be formalized without impredicativity.
Second, diverging elements are not represented.
We can apply a partial function only to the elements in its domain, but we cannot translate an expression $(f\ a)$ if $a$ does not satisfy $D$.

I propose a different approach.
The type of partial functions from $A$ to $B$ is defined as $A\rightarrow \copar{B}$, where $\copar{B}$ is a type in which $B$ can be embedded and that contains terms representing partial elements.
The standard choice in the semantics of programming languages is to put $\copar{B}\equiv B+\{\bot\}$, where $\{\bot\}$ is a one-element set \cite{scott:1982,winskel:1993}.
This solution does not obtain in type theory.
Sum types are decidable, that is, it is decidable whether an element of $B+\{\bot\}$ is in $B$ or is $\bot$.
This would imply that we can decide when a function diverges.
It is then clear that not all recursive functions are representable.
However, adopting ideas of Scott \cite{scott:1982} we can constructivize this notion by formalizing finite elements and then take the filter completion.
This approach was adopted by Michael Hedberg \cite{hedberg:1996}, who formalized constructive domain theory as a step toward a model of partial type theory inside total type theory.

Instead, here we choose as $\copar{B}$ a certain coinductive type.
Coinductive types were first introduced in type theory by Hagino \cite{hagino:1987}, their second order implementation was studied by Wraith \cite{wraith:1989} and Geuvers \cite{geuvers:1992}.
The version we use here was developed by Coquand \cite{coquand:1993} and implemented in Coq by Gimenez \cite{gimenez:1994}.

Before defining $\copar{B}$ formally, let us see the model of computation on which it is based.

We give a description of computable functions that is Turing complete and is as simple as standard formalizations like Kleene's recursive functions, the pure $\lambda$-calculus, and Turing machines.
It has the advantage that it translates easily in type theory.

We start with a system $\lambalg$ containing algebraic data types with definition by cases and function types with $\lambda$-abstraction and application.
Then we introduce a new rule $\itco$ to define functions.
To define a function $f$ from a type $A$ to a type $B$ we assign to every element $x$ of $A$ either directly a value $b$ in $B$, and we indicate it by $\return{b}$, or another element $a'$ in $A$, and we indicate it by $\step{a'}$.
In other words, we give a function $g\colon A\rightarrow A+B$ and we indicate by $\step{\cdot}$ and $\return{\cdot}$ the left and right injection, respectively.
The rule $\itco$ states that we obtain a function $f=(\itcompute\ g)\colon A\rightarrow B:$
$$
\begin{array}{l}
f\colon A\rightarrow B\\
a_1 \mapsto \return{b} \\
a_2 \mapsto \step{a'}
\end{array}
\quad \mbox{ means }\quad
\begin{array}{l}
g\colon A\rightarrow A+B\\
a_1 = \inr\ b \\
a_2 = \inl\ a'
\end{array}
\quad
\frac{g\colon A\rightarrow A+B}{f=(\itcompute\ g)\colon A\rightarrow B}.
$$

The function $f$ has the following computation behavior.
The computation of $f$ on input $a$ proceed recursively: If $(g\ a)=\inr\ b$, then the computation of $(f\ a)$ terminates with result $b$; if $(g\ a)=\inl\ a'$, we proceed to the computation of $(f\ a')$ and let $(f\ a)\leadsto (f\ a')$: if, continuing in the process, we eventually obtain a result $(f\ a')= b$, then we put $(f\ a)= b$; on the other hand, if the evaluation always gives a new element of $A$ on which to compute $f$, then the computation will continue indefinitely and the value of $(f\ a)$ will be undetermined:
$$
\begin{array}{l}
f\ a_1 \leadsto b\\
f\ a_2 \leadsto f\ a'.
\end{array}
$$

\begin{thm}
The system $\lambalg+\itco$ is Turing-complete.
\end{thm}

\proof 
We give just an outline of the proof.
A complete version of it, formalized in type theory, is given later (Definition \ref{def:recfunctions} and Theorem \ref{th:recursion}).
We show that all recursive functions can be programmed.
The zero constant, successor, and projections can be defined directly by constructors and cases.

Let $f$ be defined by primitive recursion, that is, given $g\colon \nat^n\rightarrow \nat$ and $h\colon \nat^{n+2}\rightarrow \nat$, we have
$$
\begin{array}{l}
f\colon \nat^{n+1}\rightarrow \nat\\
(f\ \seq{x}\ 0) = (g\ \seq{x})\\
(f\ \seq{x}\ (y+1)) = (h\ \seq{x}\ y\ (f\ \seq{x}\ y)).
\end{array}
$$
We use a technique called {\em continuation-passing style programming\/}, that consists in seeing programs as maps on functions \cite{reynolds:1993}.
We add a functional parameter and define a function on $\nat^n\times (\nat\rightarrow\nat)\times\nat$ using definition by cases and the rule $\itco$:
$$
\begin{array}{l}
\langle \seq{x},u,0\rangle \mapsto \return{(u\ (g\ \seq{x}))}\\
\langle \seq{x},u,y+1\rangle \mapsto
  \step{\langle \seq{x},
                \lambda z.(u\ (h\ \seq{x}\ y\ z)),y
        \rangle}.
\end{array}
$$
Call $f^+$ the function so defined, $f^+\colon \nat^n\times (\nat\rightarrow\nat)\times\nat \rightarrow \nat$.
Then we put
$$
\begin{array}{l}
f\colon \nat^{n+1} \rightarrow \nat\\
f\ \seq{x}\ y = f^+\ \langle \seq{x}, \id, y \rangle.
\end{array}
$$

Let $f$ be defined by minimization, that is, given $g\colon \nat^{n+1}\rightarrow \nat$, we have
$$
\begin{array}{l}
f\colon \nat^n\rightarrow \nat\\
f\ \seq{x} = \minimize\ y.(g\ \seq{x}\ y)
\end{array}
$$
where $\minimize\ y.(g\ \seq{x}\ y)$ denotes the least natural number $y$ for which $(g\ \seq{x}\ y)=0$ if this value exists, it is undefined otherwise.
Also in this case we use the technique of {\em accumulating results\/} by adding a new natural number parameter containing the partial result.
We use $\itco$ to define a function $f^+$ on $\nat^{n+1}$.
$$
\langle \seq{x}, y \rangle \mapsto
\begin{cases}
\return{y} & \text{if $(g\ \seq{x}\ y)=0$}\\
\step{\langle \seq{x}, y+1\rangle} & \text{otherwise.}
\end{cases}
$$
This defines a function $f^+\colon \nat^{n+1}\rightarrow \nat$.
Now we just put
$$
\begin{array}{l}
f\colon \nat^n\rightarrow \nat\\
f\ \seq{x} = (f^+\ \seq{x}\ 0).
\end{array}
$$
\qed

The definition of a recursive function $f\colon A\rightarrow B$ is then given by a map from $A$ to $A+B$, if we write $\step{a}$ and $\return{b}$ for the left and right injections, respectively.
Such a function is a coalgebra for the functor $F_BX := X+B$.
(See \cite{jacobs/rutten:1997} and \cite{aczel:2002} for elementary introductions to coalgebras.)
We recall the categorical notion of coalgebra: A coalgebra for an endofunctor $F$ is an object $A$ together with a morphism $u \colon A\rightarrow FA$.
If the functor $F$ has a final coalgebra $\nu F$, $\gamma:\nu F \rightarrow F\nu F$, then all coalgebras can be embedded in it.
This means that if $A$ and $u$ constitute an $F$-coalgebra, then there is a unique morphism $\hat{u}\colon A\rightarrow \nu F$ such that $\gamma\circ\hat{u} = F\hat{u}\circ u$.
In type theory, final coalgebras are modeled by coinductive types.
Let then $\copar{B}$ be the coinductive type associated to $F_B$.
Then, as we have said, every coalgebra $u\colon A\rightarrow A+B$ defines a function $\hat{u}\colon A \rightarrow \copar{B}$.
In conclusion, we can model partial recursive functions from a type $A$ to a type $B$ in type theory by elements of $A\rightarrow \copar{B}$.

The rest of the paper formalizes this idea and compares it with other formalizations of partial recursive functions.

Section \ref{sec:type_theory} gives a succinct introduction to type theory, with special emphasis on recursive (inductive and coinductive) types.
Section \ref{sec:partial} presents the type constructor for partial elements, the formal definition of convergence and divergence, and the formal proofs of basic properties.
Section \ref{sec:recursion} contains the representation of general recursive functions in type theory using the type for partial elements.
Section \ref{sec:nested} tackles the insidious problem of the representation of nested recursive functions.
Section \ref{sec:fixpoints} shows how to construct least fixed points of function operators.
Section \ref{sec:lazy} discusses a variant of the construct that supports a lazy evaluation order.
Finally, Section \ref{sec:monad} proves that the constructor for partial elements is a strong monad.

The whole development has been completely formalized using the proof assistant Coq \cite{coq}, version 8.
Specifically, every numbered definition has been formalized and all numbered lemmas and theorems have been proved formally; with the exception of the content of Section \ref{sec:recursion}, that we can see as a consequence of the the more abstract results of Section \ref{sec:fixpoints}.
The main results, both in this article and in the Coq development, are Theorems \ref{thm:fixed_point} and \ref{thm:least_prefix}.
The examples of nested recursive functions of Sections \ref{sec:nested} have also been programmed in Coq.
The file of the formalization is available on the author's web site:

\href{http://www.science.uottawa.ca/~vcapr396/}
     {\tt http://www.science.uottawa.ca/\~{}vcapr396/}

\section{Type Theory and Recursive Types}\label{sec:type_theory}
We work in a dependent type theory with inductive and coinductive types.
If you are familiar with this kind of system, you can skip this section.
See \cite{barendregt:1992}, \cite{barendregt/geuvers:2001}, or \cite{barthe/coquand:2002} for a good introduction to dependent type theory.
We work in a system $\lambdacoind$ consisting of the the pure type system $\lambdaP$ with the addition of sum types and (co)inductive types.
We use a formulation of (co)inductive types as recursive sum types, as in \cite{gimenez:1994}.

We use some simplifying notation to make the treatment more intuitive and concise.
We write both $\type$ and $\prop$ for the sort of small types, usually denoted by $*$ in $\lambdaP$, to distinguish computational from logical types and to facilitate applications to other type systems in which these sorts are distinguished.
If $\Gamma$ is a valid context and $\Delta$ is a sequence of assignments of types to variables, we write $\Gamma\vdash \valcont{\Delta}$ to express that $\Gamma,\Delta$ is also a valid context.
We denote by $\seq{x_\Delta}$ the sequence of variables assumed in $\Delta$.

If $\Delta \equiv x_1\colon A_1, \ldots, x_m\colon A_m$ and $\seq{d}\equiv d_1, \ldots, d_m$ is a sequence of pseudo-terms, then the judgment $\Gamma \vdash \seq{d}\ccolon \Delta$ expresses the conjunction of the judgments
$$
\begin{array}{l}
\Gamma\vdash d_1\colon A_1,\\
\Gamma\vdash d_2\colon A_2[d_1/x_1],\\
\vdots \\
\Gamma\vdash d_m\colon A_m[d_1/x_1, \ldots, d_{m-1}/x_{m-1}].
\end{array}
$$
If $\Gamma, \Delta\vdash t\colon T$ and $\Gamma \vdash \seq{d}\ccolon \Delta$, we write $t[\seq{d}]$ for $t[d_1/x_1,\ldots,d_m/x_m]$, leaving it implicit that we are substituting the last $n$ variables in the context of $t$.
To avoid confusion, we may sometimes introduce the term $t$ as $t[x_1,\ldots,x_n]$..

If $\Gamma,\Delta\vdash B:\type$, we write $\Pi\Delta.B$ for the successive product of $B$ over each of the variables in $\Delta$: $\Pi\Delta.B \equiv \Pi x_1\colon A_1.\Pi x_2\colon A_2 \ldots \Pi x_n\colon A_n.B$. 
We also use the notation $(\Delta)B$ for $\Pi\Delta.B$.

We add constant definition to the system, that is, if $\Gamma\vdash e\colon T$, then we can add a declaration of the form $\ttf{t}:=e$ to the context, with the typing rule $\Gamma,\ttf{t}:=e\vdash \ttf{t}\colon T$ and the reduction rule $\ttf{t} \leadsto_\delta e$.
A constant name can be declared at most once.
The classes of names for variables and constants are disjoint.
However, sometimes we use the same name in different fonts to denote a variable and a constant.
It will then be understood that the variable is automatically substituted with the constant whenever the latter is defined in the context.
For example, if we have $\Gamma,t\colon T\vdash u\colon U$ and $\Gamma \vdash e\colon T$, we write $\Gamma,\ttf{t}:=e\vdash u\colon U$ for $\Gamma,\ttf{t}:=e\vdash u[\ttf{t}/t]\colon U$.

Sum types are specified by a sequence of constructors with arguments of previously defined types.

\begin{defi}
Let $\Gamma$ be a valid context and assume that the following judgments are derivable:
$$
\begin{array}{l}
\Gamma\vdash \valcont{\Delta},\\
\Gamma\vdash \valcont{\Theta_i},\\
\Gamma, \Theta_i \vdash \seq{p_i}\ccolon \Delta
\end{array}
$$
for $1\leq i \leq n$.
Let $\ttf{T}$ and $\ttf{c}_1$, \ldots, $\ttf{c}_n$ be new constant names.
Then we can add the definition of the sum type
$$
\begin{sumtype}{\ttf{T}}{[\Delta]\colon \type}
\ttf{c}_1 \colon (\Theta_1)(\ttf{T}\ \seq{p_1})\\
\vdots \\
\ttf{c}_n \colon (\Theta_n)(\ttf{T}\ \seq{p_n})
\end{sumtype}
$$
in the context $\Gamma$.
Let $\Gamma'$ be the extension of $\Gamma$ with this sum type definition.
We have the following rules for $\ttf{T}$:
\begin{rules}
\item[Formation]
$$
\frac{\Gamma'\vdash \seq{d}\ccolon \Delta}
     {\Gamma'\vdash (\ttf{T}\ \seq{d})\colon \type};
$$
\item[Introduction]
$$
\frac{\Gamma'\vdash \seq{a}\ccolon \Theta_i}
     {\Gamma'\vdash (\ttf{c}_i\ \seq{a})\colon (\ttf{T}\ \seq{p_i}[\seq{a}])}
\quad\mbox{for }1\leq i\leq n;
$$
\item[Elimination (definition by cases)]
$$
\frac{\begin{array}{l}
      \Gamma'\vdash P\colon (\Delta)(\ttf{T}\ \seq{x_\Delta})
                            \rightarrow \type\\
      \Gamma',\Theta_i\vdash e_i\colon (P\ \seq{p_i}\ (\ttf{c}_i\ \seq{x_{\Theta_i}}))
        \quad 1\leq i\leq n\\
      \Gamma'\vdash \seq{d}\ccolon \Delta\\
      \Gamma'\vdash t\colon (\ttf{T}\ \seq{d})
      \end{array}
     }
     {\Gamma'\vdash \left(
                    \begin{bycases}{t}
                    (\ttf{c}_1\ \seq{x_{\Theta_1}}) & e_1\\
                    \multicolumn{2}{l}{\vdots}\\
                    (\ttf{c}_n\ \seq{x_{\Theta_n}}) & e_n
                    \end{bycases}
                    \right)
                    \colon (P\ \seq{d}\ t)
     };
$$
\item[Reduction]
$$
\left(
\begin{bycases}{(\ttf{c}_i\ \seq{a})}
(\ttf{c}_1\ \seq{x_{\Theta_1}}) & e_1\\
\multicolumn{2}{l}{\vdots}\\
(\ttf{c}_n\ \seq{x_{\Theta_n}}) & e_n
\end{bycases}
\right)
\reduce
e_i[\seq{a}/\seq{x_{\Theta_i}}]
\quad\mbox{for }1\leq i\leq n.
$$
\end{rules}
\end{defi}

When a function is defined by case analysis, we use a {\em pattern matching\/} notation to facilitate reading.
A function
$$
\begin{array}{l}
\ttf{f}\colon (\Delta)(t\colon (\ttf{T}\ \seq{x_\Delta}))
              (P\ \seq{x_\Delta}\ t)\\
\ttf{f} \equiv  \lambda\seq{x_\Delta}\ccolon \Delta. \lambda t\colon (\ttf{T}\ \seq{x_\Delta}).
                \begin{bycases}{t}
                (\ttf{c}_1\ \seq{x_{\Theta_1}}) & e_1\\
                \multicolumn{2}{l}{\vdots}\\
                (\ttf{c}_n\ \seq{x_{\Theta_n}}) & e_n
                \end{bycases}
\end{array}
$$
is written
$$
\begin{array}{l}
\ttf{f}\colon (\Delta)(t\colon (\ttf{T}\ \seq{x_\Delta}))
              (P\ \seq{x_\Delta}\ t)\\
\ttf{f}\ \seq{p_1}\ (\ttf{c}_1\ \seq{x_{\Theta_1}}) = e_1 \\
\vdots\\
\ttf{f}\ \seq{p_n}\ (\ttf{c}_n\ \seq{x_{\Theta_n}}) = e_n.
\end{array}
$$

We formulate both inductive and coinductive types as recursive sum types.
By this we mean that the type $\ttf{T}$ may occur in the types of arguments of the constructors $\Theta_i$.
The only restriction is that these occurrences must be strictly positive.
See \cite{coquand/paulin:1990,gimenez:1994} for the notion of strictly positive occurrence.

Inductive and coinductive types were introduced in typed $\lambda$-calculus by Hagino in \cite{hagino:1987a,hagino:1987} with the name of {\em categorical data types\/}.
In \cite{wraith:1989} and \cite{geuvers:1992} their expression in polymorphic typed $\lambda$-calculus was studied.

In our formulation, inductive and coinductive types have the same rules as sum types plus a fixpoint rule for inductive types and a cofixpoint rule for coinductive types allowing the definition of recursive functions.

Inductive types were introduced in dependent type theory in \cite{coquand/paulin:1990} (see also \cite{pfenning/paulin:1990}, \cite{paulin:1993}, and \cite{werner:1994}).
Coinductive types were studied in \cite{mpc:1986,mendler:1987,coquand:1993}.
The version given here comes from \cite{gimenez:1994} and is the one implemented in the proof tool Coq \cite{coq:2002}.

\begin{defi}
We make the same assumptions as in the definition of sum types, except that
$$
\Gamma, T\colon (\Delta)\type \vdash \valcont{\Theta_i}
\qquad
\Gamma, T\colon (\Delta)\type, \Theta_i \vdash \seq{p_i}\ccolon \Delta
$$
with $T$ occurring only strictly positively in $\Theta_i$, for $1\leq i \leq n$ (see \cite{gimenez:1994}, Section 2.2, pg.43, for the definition of strict positivity).
Let $\ttf{T}$ and $\ttf{c}_1$, \ldots, $\ttf{c}_n$ be new constant names.
Then we can add the definition of the inductive type
$$
\begin{inductive}{\ttf{T}}{[\Delta]\colon \type}
\ttf{c}_1 \colon (\Theta_1)(\ttf{T}\ \seq{p_1})\\
\vdots \\
\ttf{c}_n \colon (\Theta_n)(\ttf{T}\ \seq{p_n})
\end{inductive}
$$
in the context $\Gamma$.
Let $\Gamma'$ be the extension of $\Gamma$ with this inductive type definition.
We have the same rules as for sum types, plus the following:
\begin{rules}
\item[Fixpoint]
$$
\frac{\begin{array}{l}
      \Gamma' \vdash P\colon (\Delta)(\ttf{T}\ \seq{x_\Delta})
                             \rightarrow \type\\
      \Gamma', f\colon (\Delta)(t\colon (\ttf{T}\ \seq{x_\Delta}))
                       (P\ \seq{x_\Delta}\ t),
               \Delta,t\colon (\ttf{T}\ \seq{x_\Delta})
               \vdash e\colon (P\ \seq{x_\Delta}\ t)
      \end{array}
     }
     {\Gamma'\vdash (\fix\ [f,\seq{x_\Delta},t]e)\colon
                    (\Delta)(t\colon (\ttf{T}\ \seq{x_\Delta}))
                            (P\ \seq{x_\Delta}\ t)
     }
\mathcal{D}\{f,t,e\}
$$
where $\mathcal{D}\{f,t,e\}$ is a side condition defined in \cite{gimenez:1994}, Section 3.1, pg. 47;
\item[Reduction]
$$
(\ttf{f}\ \seq{d}\ (\ttf{c}_i\ \seq{a})) \reduce
e[\ttf{f},\seq{d},(\ttf{c}_i\ \seq{a})]
$$
where $\ttf{f}\equiv (\fix\ [f,\seq{x_\Delta},t]e)$.
\end{rules}

\end{defi}

\begin{defi}
We make the same assumptions as in the definition of inductive types.
Let $\ttf{T}$ and $\ttf{c}_1$, \ldots, $\ttf{c}_n$ be new constant names.
Then we can add the definition of the coinductive type
$$
\begin{coinductive}{\ttf{T}}{[\Delta]\colon \type}
\ttf{c}_1 \colon (\Theta_1)(\ttf{T}\ \seq{p_1})\\
\vdots \\
\ttf{c}_n \colon (\Theta_n)(\ttf{T}\ \seq{p_n})
\end{coinductive}
$$
in the context $\Gamma$.
Let $\Gamma'$ be the extension of $\Gamma$ with this coinductive type definition.
We have the same rules as for sum types, plus the following, where $\Xi$ and $\Gamma''$ are valid context extensions of $\Gamma'$:
\begin{rules}
\item[Cofixpoint]
$$
\frac{\begin{array}{l}
      \Gamma',\Xi\vdash \seq{u}\ccolon \Delta \quad
      \Gamma',f\colon (\Xi)(\ttf{T}\ \seq{u})
        \vdash e\colon (\Xi)(\ttf{T}\ \seq{u})
      \end{array}
     }
     {\Gamma'\vdash (\cofix\ [f]e)\colon (\Xi)(\ttf{T}\ \seq{u})}
\mathcal{C}\{f,e\}
$$
where $\mathcal{C}\{f,e\}$ is a side condition defined in \cite{gimenez:1994}, Section 4.1, pg. 53;
\item[Reduction]
if $\Gamma',\Gamma''\vdash \seq{v}\ccolon \Xi$, then
$$
\left(
\begin{bycases}{(\ttf{f}\ \seq{v})}
(\ttf{c}_1\ \seq{x_{\Theta_1}}) & e_1\\
\multicolumn{2}{l}{\vdots}\\
(\ttf{c}_n\ \seq{x_{\Theta_n}}) & e_n
\end{bycases}
\right)
\reduce
\left(
\begin{bycases}{(e[\ttf{f}]\ \seq{v})}
(\ttf{c}_1\ \seq{x_{\Theta_1}}) & e_1\\
\multicolumn{2}{l}{\vdots}\\
(\ttf{c}_n\ \seq{x_{\Theta_n}}) & e_n
\end{bycases}
\right)
$$
where $\ttf{f}\equiv (\cofix\ [f]e)$.
\end{rules}
\end{defi}

We refer to \cite{gimenez:1994} for the definition of the side conditions $\mathcal{D}$ and $\mathcal{C}$ of the fixpoint and cofixpoint rules.
Let us give only an intuitive reminder of their meaning.
The side condition $\mathcal{D}\{f,t,e\}$ guarantees that whenever the fixpoint function is applied to an argument in the inductive type, it performs recursive calls only on structurally smaller objects.
The side condition $\mathcal{C}\{f,e\}$ guarantees that whenever the cofixpoint function is applied to obtain an element of the coinductive type, this element is {\em productive\/} \cite{coquand:1993}, that is, it can be reduced to constructor form.

We will often use a simpler notation for fixed and cofixed points.
We write $\ttf{f} = e[\ttf{f}]$ for $\ttf{f}\equiv (\fix\ [f,\seq{x_\Delta},t]e)$ or $\ttf{f}\equiv (\cofix\ [f]e)$.
Which one of the two is meant will be clear from the type.
There can be ambiguity when the domain of $\ttf{f}$ is an inductive type and its codomain is a coinductive type.
In that case we state explicitly whether a fixed or cofixed point is meant.

When coinductive types are considered categorically as the dual of inductive types, their formulation is different from the one given here.
They are not recursive sum types like the inductive ones, but rather recursive record types, since records are the dual of sums.
This is the way they are presented in \cite{geuvers:1992}.
Our formulation conforms to the intuitive conception of coinductive types as types of possibly infinite elements.
Objects of coinductive types are built up by constructors as those of inductive types, but while inductive objects must be wellfounded, coinductive ones may be infinitely deep.
This formulation is given in \cite{coquand:1993}.
The equivalence of the two formulations is proved in \cite{gimenez:1994}.

\section{Coinductive types of partial elements}\label{sec:partial}
For every type $A$, we define a coinductive type whose elements can be thought of as possibly undefined elements of $A$.
\begin{defi}
Let $A$ be a type, that is, $\Gamma \vdash A\colon \type$.
Then we define
$$
\begin{coinductive}{\copar{A}}{\colon\type}
\returnsym\colon A\rightarrow \copar{A}\\
\stepsym \colon \copar{A}\rightarrow \copar{A}.
\end{coinductive}
$$
We use the notation $\return{a}$ for $(\returnsym\ a)$ and $\step{x}$ for $(\stepsym\ x)$.
\end{defi}

Intuitively, an element of $\copar{A}$ is either an element of $A$ or a computation step followed by an element of $\copar{A}$.
Since in coinductive types it is possible to define infinite elements,
there is an object $\step{\step{\step{\cdots}}}$ denoting an infinite computation.
Formally, it is defined as
$$
\stepfun^\infty = \step{\stepfun^\infty}\colon \copar{A}.
$$

We want to identify all terms of the form $\mathop{\stepfun^k}\return{a}$ as equivalent representations of the element $a$.
We do it by defining an equality relation on $\copar{A}$ that is a strengthening of bisimulation (see, for example, \cite{milner:1980,aczel:1988,barwise/moss:1996,hermida/jacobs:1998}).
First we define, inductively, when an element of $\copar{A}$ converges to a value in $A$ and, coinductively, when it diverges.
\begin{defi}
$$
\begin{array}{l}
\begin{biginductive}{\parvaluesym}{[x\colon \copar{A}, a\colon A]\colon \prop}
\parvalreturn \colon (a\colon A)(\parvaluesym\ \return{a}\ a)\\
\parvalstep   \colon (x\colon \copar{A}; a\colon A)(\parvaluesym\ x\ a)
                     \rightarrow (\parvaluesym\ \step{x}\ a)
\end{biginductive}\\
\\
\begin{bigcoinductive}{\divergesym}{[x\colon \copar{A}]\colon \prop}
\divergecons \colon (x\colon \copar{A})(\divergesym\ x)
                    \rightarrow (\divergesym\ \step{x})
\end{bigcoinductive}
\end{array}
$$
We use the notation $\parvalue{x}{a}$ for $(\parvaluesym\ x\ a)$ and $\diverge{x}$ for $(\divergesym\ x)$.
\end{defi}

Our first example of a proof by coinduction shows that $\stepfun^\infty$ diverges.
This is trivial, but we give the proof to illustrate the style of derivation that is used to prove coinductive predicates.
\begin{lem}\label{infinite_diverge}
$
\diverge{(\stepfun^\infty)}.
$
\end{lem}
\proof 
If we think of coinductive definitions as representing infinite objects, coinductive predicates are proved by infinite proofs.
The infinite proof showing the truth of the statement is
$$
H = (\divergecons\ \stepfun^\infty\ (\divergecons\ \stepfun^\infty\ \cdots ))
\colon \diverge{(\stepfun^\infty)}.
$$
It consists of an infinite sequence of nested applications of the constructor $\divergecons$, always with $\stepfun^\infty$ as first argument.
At first sight this may seem a circular proof that doesn't prove anything.
However, if we analyze it, we discover that it actually establishes what it alleges.
Let us give names to the subterms of $x_0=\stepfun^\infty$:
$$
x_0=\step{x_1} \quad x_1=\step{x_2} \quad x_2=\step{x_3} \quad \cdots.
$$
Of course, they all coincide:
$x_0=x_1=x_2=x_3=\cdots = \stepfun^\infty.$
Let us call $H_i$ the proof of $\diverge{x_i}$.
Again, it will be clear that we can use the same proof for every $H_i$.
The first step of the proof constructs $H_0$ from $H_1$:
$$
H_0 = (\divergecons\ x_1\ H_1).
$$
This just says that if $x_1$ diverges, then $x_0$ must also diverge because it is obtained by adding one $\stepfun$ to $x_1$.
In turn the proof $H_1$ is constructed by
$$
H_1 = (\divergecons\ x_2\ H_2)
$$
and so on, producing the infinite proof, which now doesn't seem purposeless anymore: Every occurrence of $\divergecons$ shows that the term contains a separate $\stepfun$ constructor.

In practice, we will define this proof $H$ in terms of itself:
$$
H = (\divergecons\ \stepfun^\infty\ H).
$$
Formally, this definition is valid because the recursive occurrence of $H$ is an argument of $\divergecons$, that is, in coinductive jargon, $H$ is {\em guarded\/} by $\divergecons$.
This guarantees that $H$ can be expanded to an infinite proof not containing any occurrence of $H$ itself.
\qed

Intuitively, $\copar{A}$ contains copies of the elements of $A$ plus a diverging element $\stepfun^\infty$.
However, $\copar{A}$ is not isomorphic to $A+\onet$, where $\onet$ is the one element type, because divergence is undecidable.
In fact, the proposition
$$
\forall x\colon \copar{A}.
(\exists a\colon A. \parvalue{x}{a}) \lor \diverge{x}
$$
is not provable, while the corresponding
$$
\forall x\colon A+\onet.
(\exists a\colon A.x=(\inl\ a)) \lor x=(\inr\ 0_\onet)
$$
is trivially true by case analysis.
However, a different property, classically equivalent to the above decidability property, is constructively provable.
\begin{thm}\label{not_value_diverge}
$$
\forall x\colon \copar{A}.(\neg \exists a\colon A.\parvalue{x}{a})\rightarrow \diverge{x}.
$$
\end{thm}
\proof 
This is our first non-trivial proof by coinduction, so we take care of doing it in detail and explain how such proofs work.
We know that a coinductive object can be defined in terms of itself, as long as the definition is {\em productive,\/} that is, it can generate leading constructors to an arbitrary depth.
This is true for coinductive proofs as well, leading to the apparently paradoxical fact that we can assume the statement that we want to prove, provided that we use it in such a way that every step of the (infinite) proof can be generated.

Let us see how this works specifically in our case.
We give the name $H$ to the proof that we are constructing:
$$
H\colon \forall x\colon \copar{A}.(\neg \exists a\colon A.\parvalue{x}{a})\rightarrow \diverge{x}.
$$
We now describe how this function is constructed.
Let $x$ be an element of $\copar{A}$.
By case analysis we know that either $x=\return{a}$ for some $a\colon A$ or $x=\step{x'}$ for some other $x'\colon \copar{A}.$
We consider separately the two cases.

If $x=\return{a}$, then trivially $\exists a\colon A.\parvalue{x}{a}$.
The premise of the statement is then false and, therefore, the implication is trivially true.

If, on the other hand, $x=\step{x'}$, let us assume that there is no $a\colon A$ such that $\parvalue{x}{a}$. We have to prove that $x$ diverges.
The constructor $\divergecons$ allows us to deduce this, if we can prove that $x'$ diverges.
Now we apply the proof $H$ to $x'$: If we can prove $\neg \exists a\colon A.\parvalue{x'}{a}$, then we can conclude that $x'$ diverges.
But this assertion is true because, if there were an $a$ such that $\parvalue{x'}{a}$, then, by $\parvalstep$, also $\parvalue{x}{a}$ against the assumption.

In conclusion, $H$ allows us to deduce that $x'$ diverges and so, by $\divergecons$, that $x$ diverges too, as desired.
We used the proof $H$ inside the definition of $H$ itself; isn't this circular?
No, because we used it to prove an hypothesis ($\diverge{x'}$) that was generated by $\divergecons$.
In the coinduction jargon, the occurrence of $H$ is {\em guarded\/} by the occurrence of $\divergecons$.

If you want to convince yourself of the soundness of this proof style, consider what concrete proofs are generated by $H$ in specific cases.
For a converging element $x = \step{\step{\cdots \step{\return{a}}}}$, the proof is $(H\ x)$.
Let us expand it.
Since $x = \step{x'}$ for $x' = \step{\cdots \step{\return{a}}}$, the second case of the proof applies:
$$
(H\ x) = (\divergecons\ x'\ (H\ x')).
$$
In turn, the proof $(H\ x')$ reduces similarly, and so on, until we reach the proof
$$
(H\ x) = (\divergecons\ x'\ (\divergecons\ x''\ \cdots (\divergecons\ \return{a}\ (H\ \return{a}))\cdots)).
$$
Finally, $(H\ \return{a})$ trivially reduces to a proof by {\em ex falso quodlibet\/} as described in the first case, in which $H$ does not occur.

On the other hand, if $x = \stepfun^\infty$, the described reduction procedure of $(H\ x)$ continues indefinitely, generating the infinite proof
$$
(H\ \stepfun^\infty) = (\divergecons\ \stepfun^\infty\ (\divergecons\ \stepfun^\infty\ \cdots ))
$$
of Lemma \ref{infinite_diverge}.
This proof also does not contain any occurrence of $H$.
\qed

From now on, proofs of coinductive properties are less prolix.
We take it for granted that we can assume the validity of the statement that we are proving.
In case of doubt, it can always be checked, intuitively, that the proofs are productive or, formally, that they satisfy the condition $\mathcal{C}\{f,e\}.$
The correctness of the proofs has been checked formally using the proof assistant Coq.

Now all the elements are in place for the definition of equality of partial elements:
It is also a coinductive relation.
\begin{defi}
Let $A$ be a type.
We define equality on $\copar{A}$ by
$$
\begin{bigcoinductive}{\coeqrel_A}{[x,y\colon \copar{A}]\colon \prop}
\coeqvalue\colon (x,y\colon \copar{A}; a\colon A)
                 \parvalue{x}{a} \rightarrow \parvalue{y}{a}
                 \rightarrow (\coeqrel_A\ x\ y)\\
\coeqstep\colon (x,y\colon \copar{A})
                (\coeqrel_A\ x\ y) \rightarrow
                (\coeqrel_A\ \step{x}\ \step{y})
\end{bigcoinductive}
$$
We use the notation $x\coeq y$ for $(\coeqrel_A\ x\ y)$.
\end{defi}

Note that, since convergence and divergence are in general undecidable, the equality $x\coeq y$ is not equivalent to the proposition
$$
(\exists a\colon A.\parvalue{x}{a} \land \parvalue{y}{a}) \lor
(\diverge{x} \land \diverge{y}).
$$

A series of results can be proved easily, stating that the defined operations, relations, and predicates behave as expected.
\begin{lem}\label{partial_basic}
For all $x,y\colon \copar{A}$ and $a,a_1,a_2\colon A$, we have
$$
\begin{array}{rl@{\qquad}rl}
1.&  \parvalue{x}{a} \rightarrow \parvalue{y}{a} \rightarrow x\coeq y &
2.&  \diverge{x} \rightarrow \diverge{y} \rightarrow x\coeq y \\[1ex]
3.&  \parvalue{x}{a} \rightarrow x\coeq y \rightarrow \parvalue{y}{a} &
4.&  \diverge{x} \rightarrow x\coeq y \rightarrow \diverge{y} \\[1ex]
5.&  \parvalue{x}{a} \rightarrow x\coeq\return{a} &
6.&  x\coeq\return{a} \rightarrow \parvalue{x}{a} \\[1ex]
7.&  \parvalue{\step{x}}{a} \rightarrow \parvalue{x}{a} &
8.&  \diverge{\step{x}} \rightarrow \diverge{x} \\[1ex]
9.&  \parvalue{x}{a} \rightarrow \neg{\diverge{x}} &
10.& \parvalue{x}{a} \rightarrow \diverge{y}\rightarrow \neg x\coeq y \\[1ex]
11.& x \coeq y \rightarrow x \coeq \step{y} &
12.& \step{x} \coeq y \rightarrow x \coeq y \\[1ex]
13.& \parvalue{\return{a_1}}{a_2} \rightarrow a_1=a_2 &
14.& x \coeq x \\[1ex]
15.& x \coeq y \rightarrow y \coeq x &
16.& x \coeq y \rightarrow y \coeq z \rightarrow x \coeq z
\end{array}
$$
\end{lem}
\proof 
\begin{itemize}

\item[1.]
By $\coeqvalue.$

\item[2.]
Proof by coinduction.
Assume $H$ is the proof of the statement:
$$
H\colon \forall x,y:\copar{A}.\diverge{x} \rightarrow \diverge{y} \rightarrow x\coeq y.
$$
Let $x$ and $y$ be diverging elements.
Then they must be in the form $x=\step{x'}$ and $y=\step{y'}$, where $x'$ and $y'$ are also diverging.
We apply the proof $H$ to $x'$ and $y'$ to obtain $x'\coeq y'$.
By $\coeqstep$ we also have $x\coeq y$.
The recursive application of $H$ is guarded by $\coeqstep$.

\item[5.]
Assume $\parvalue{x}{a}$.
By $\parvalreturn$ we have that $\parvalue{\return{a}}{a}$.
Thus, by $\coeqvalue$ we can conclude that $x\coeq \return{a}$.

\item[13.]
Assume that $\parvalue{\return{a_1}}{a_2}$.
Any proof $H$ of this statement must be constructed by using $\parvalreturn$, since the only other constructor $\parvalstep$ gives a conclusion of the wrong form, that is, with a $\stepfun$ constructor in the first argument.
Therefore, it must be
$$
H = (\parvalreturn\ a')
$$
for some $a'$.
But, since $(\parvalreturn\ a')$ is a proof of $\parvalue{\return{a'}}{a'}$, we may conclude that $a'=a_1=a_2$.
This is an example of {\em proof by inversion\/}; it consists in proving a goal by reasoning on the possible form of an assumption. 

\item[6.]
Assume that $x\coeq\return{a}$.
We proceed again by inversion on this assumption.
If $H$ is a proof of this premise, it must have been obtained by an application of $\coeqvalue$, since the other constructor $\coeqstep$ gives a conclusion with the wrong form, that is, with a $\stepfun$ constructor in the second argument.
Therefore, it must be
$$
H = (\coeqvalue\ x'\ y'\ a'\ h_1\ h_2)
$$
for some $x',y'\colon \copar{A}$, $a'\colon A$, $h_1$ a proof of $\parvalue{x'}{a'}$, and $h_2$ a proof of $\parvalue{y'}{a'}$.
But this is a proof of $x'\coeq y'$, so we can conclude that $x'=x$ and $y'=\return{a}$.
Then $h_2$ is a proof of $\parvalue{\return{a}}{a'}$, from which we deduce, by point 13, that $a=a'$.
But then $h_1$ is a proof of $\parvalue{x}{a}$, and the conclusion is established.

\item[14.]
By coinduction: Assume that $H$ is the proof of $\colon \forall x\colon \copar{A}. x \coeq x.$
Any element $x$ must be of one of the two forms: $x=\return{a'}$ or $x=\step{x'}$.

If $x=\return{a'}$, we must prove that $\return{a'}\coeq \return{a'}.$
By $\parvalreturn$ we have that $\parvalue{\return{a'}}{a'}$, so, by $\coeqvalue$, we have that $\return{a'}\coeq \return{a'}.$

If $x=\step{x'}$ we must prove that $\step{x'}\coeq \step{x'}$.
By $H$ we have that $x'\coeq x'$, so, by $\coeqstep$, we have the conclusion.
The application of $H$ is guarded by $\coeqstep$.

\item[15.]
By coinduction: Assume that $H$ is the proof of $\forall x,y\colon \copar{A}.x \coeq y \rightarrow y \coeq x.$
Assume $x\coeq y$ and let $h$ be a proof of this premise.
We proceed by inversion on the proof $h$.
It must have been obtained by applying one of the two constructors $\coeqvalue$ or $\coeqstep$.

If $h = (\coeqvalue\ x\ y\ a'\ h_1\ h_2)$, with $h_1$ a proof of $\parvalue{x}{a'}$ and $h_2$ a proof of $\parvalue{y}{a'};$
then we just have to apply the same constructor with inverted arguments to obtain the desired goal:
$(\coeqvalue\ y\ x\ a'\ h_2\ h_1)$ is a proof of $y\coeq x$.

If $h = (\coeqstep\ \step{x'}\ \step{y'}\ h')$, with $x=\step{x'}$, $y=\step{y'}$, and $h'$ a proof of $x'\coeq y'$; then we apply $H$ to $x', y',$ and $h'$ to obtain $y'\coeq x'$, and then we have $y'=\step{y'}\coeq \step{x'}=x$ by $\coeqstep$.
The application of $H$ is guarded by $\coeqstep$.

\item[7.]
By inversion on the proof of the premise $\parvalue{\step{x}}{a}$, which can only have been obtained by using $\parvalstep$ on a proof of $\parvalue{x}{a}$.

\item[11.]
By coinduction: Assume that $H$ is the proof of $\forall x,y\colon \copar{A}. x \coeq y \rightarrow x \coeq \step{y}$.
Assume $x\coeq y$ and let $h$ be a proof of this premise.
We proceed by inversion on $h$.

If $h=(\coeqvalue\ x\ y\ a'\ h_1\ h_2)$, with $h_1$ a proof of $\parvalue{x}{a'}$ and $h_2$ a proof of $\parvalue{y}{a'};$ then, by $\parvalstep$, from $h_2$ we can deduce $\parvalue{\step{y}}{a'}$.
By $\coeqvalue$, we conclude that $x\coeq \step{y}$.

If $h=(\coeqstep\ \step{x'}\ \step{y'}\ h')$, with $x=\step{x'}$, $y=\step{y'}$, and $h'$ a proof of $x'\coeq y'$; then we apply $H$ to $x', y',$ and $h'$ to obtain $x'\coeq \step{y'}$.
By $\coeqstep$ we then have that $x=\step{x'} \coeq \step{\step{y'}} = \step{y'}$.
The application of $H$ is guarded by $\coeqstep$.

\item[12.]
By inversion on the proof of the premise $\step{x} \coeq y$ using point 7 in case $\coeqvalue$ was used and point 11 in case $\coeqstep$ was used.

\item[3.]
Assume that $\parvalue{x}{a}$ and let $h$ be a proof of this premise.
We proceed by induction on the structure of $h$.

If $h = (\parvalreturn\ a)\colon \parvalue{\return{a}}{a}$ then we have to prove that if $\return{a}\coeq y$ then $\parvalue{y}{a}$.
By point 15, $y\coeq \return{a}$ and, by point 6, $\parvalue{y}{a}$.

If $h = (\parvalstep\ x'\ a\ h')$ with $x=\step{x'}$ and $h$ a proof of $\parvalue{x'}{a}$, we know by induction hypothesis that $\forall y\colon \copar{A}. x'\coeq y \rightarrow \parvalue{y}{a}$.
Assume now that $x\coeq y$, that is, $\step{x'}\coeq y$.
By point 12 we can deduce that $x'\coeq y$, and therefore, by induction hypothesis, that $\parvalue{y}{a}$.

\item[8.]
By inversion on the proof of the premise $\diverge{\step{x}}$.

\item[9.]
By induction on the proof $h$ of the premise $\parvalue{x}{a}$.

If $h = (\parvalreturn\ a)$, with $x=\return{a}$, we have to prove that $\neg \diverge{\return{a}}$.
This is true because it is impossible to build a proof of $\diverge{\return{a}}$.
If such a proof existed, it should be obtained by an application of the only constructor $\divergecons$ of the predicate $\divergesym$.
But such a proof $(\divergecons\ x\ h)$ can only prove statements in the form $\diverge{\step{x}}$.
Since in our case the argument of $\divergesym$ is $\return{a}$, we reach a contradiction.
This proof is a degenerate case of proof by inversion: By analyzing the possible form of the proof we conclude that no form is possible.

If $h = (\parvalstep\ x'\ a\ h')$ with $x=\step{x'}$ and $h$ a proof of $\parvalue{x'}{a}$, we know by induction hypothesis that $\neg \diverge{x'}$.
To prove that $\neg \diverge{x}$, assume that $\diverge{x}$, that is $\diverge{\step{x'}}$.
By point 8 we then have that $\diverge{x'}$, against the induction hypothesis.
Having reached a contradiction, we conclude that $\neg \diverge{x}$.

\item[4.]
By coinduction: Assume that $H$ is the proof of $\forall x,y\colon\copar{A}.\diverge{x} \rightarrow x\coeq y \rightarrow \diverge{y}$.
Assume $x\coeq y$ and let $h$ be a proof of this assumption.
We must prove that $\diverge{x}\rightarrow \diverge{y}$.
We proceed by inversion on $h$.

Suppose $h=(\coeqvalue\ x\ y\ a'\ h_1\ h_2)$ with $h_1$ a proof of $\parvalue{x}{a'}$ and $h_2$ a proof of $\parvalue{y}{a'}.$
If we assume that $\diverge{x}$ then we obtain a contradiction by point 9.
The conclusion then follows by {\em ex falso quodlibet.\/}

Suppose $h=(\coeqstep\ \step{x'}\ \step{y'}\ h')$, with $x=\step{x'}$, $y=\step{y'}$, and $h'$ a proof of $x'\coeq y'$.
From the assumption $\diverge{x}$, that is $\diverge{\step{x'}}$, and point 8 we derive  $\diverge{x'}$.
We apply $H$ to this proof and $h'$ to obtain $\diverge{y'}$.
By $\divergecons$ we then obtain $\diverge{\step{y'}}$, that is, $\diverge{y}$.
The application of $H$ is guarded by $\divergecons$.

\item[10.]
By induction on the proof $h$ of the premise $\parvalue{x}{a}$.

If $h=(\parvalreturn\ a)$, with $x=\return{a}$, we have to prove that $\diverge{y}\rightarrow \return{a}\neq y$.
Assume that $\diverge{y}$ and $\return{a}\coeq y.$
By point 15, $y\coeq \return{a}$; by point 6, $\parvalue{y}{a}$; by point 9, $\neg{\diverge{y}}$ against the hypothesis.
Discharging the second hypothesis, we obtain $\neg \return{a}\coeq y$.

If $h = (\parvalstep\ x'\ a\ h')$ with $x=\step{x'}$ and $h$ a proof of $\parvalue{x'}{a}$, we know by induction hypothesis that $\diverge{y}\rightarrow \neg x'\coeq y.$
Assume that $\diverge{y}$, so $ \neg x'\coeq y$ by induction hypothesis.
We have to prove $\neg x\coeq y$, that is, $\neg \step{x'}\coeq y$.
Assume that $\step{x'}\coeq y$.
By point 12 we then have that $x'\coeq y$, contradicting a previous conclusion.
Having reached a contradiction, we conclude that $\neg x\coeq y.$

\item[16.]
By coinduction: Assume that $H$ is the proof of $\forall x,y,z\colon\copar{A}.x \coeq y \rightarrow y \coeq z \rightarrow x \coeq z$.
Assume $h_1$ is a proof of $x\coeq y$ and $h_2$ is a proof of $y\coeq z$.
We proceed by inversion on these proofs: We have three cases.

Suppose $h_1 = (\coeqvalue\ x\ y\ a'\ k_1\ k_2)$ with $k_1$ a proof of $\parvalue{x}{a'}$ and $k_2$ a proof of $\parvalue{y}{a'}$.
By point 3, from $k_2$ and $h_2$ we can derive that $\parvalue{z}{a'}$.
By $k_1$ and $\coeqvalue$, it follows that $x\coeq z$.

Suppose $h_2 = (\coeqvalue\ y\ z\ a'\ k_2\ k_3)$. The proof is similar to the previous case.

Suppose $h_1 = (\coeqstep\ x'\ y'\ h_1')$ with $x=\step{x'}, y=\step{y'},$ and $h_1'$ a proof of $x'\coeq y';$ and $h_2 = (\coeqstep\ y'\ z'\ h_2')$ with $y=\step{y'}, z=\step{z'},$ and $h_2'$ a proof of $y'\coeq z'.$
We apply $H$ to $x',y',z'$ and the proofs $h_1'$ and $h_2',$ obtaining that $x'\coeq z'.$
By $\coeqstep$ it follows that $\step{x'}\coeq \step{z'},$ that is, $x\coeq z.$
The application of $H$ is guarded by $\coeqstep.$

\end{itemize}
\qed

Points 14, 15, and 16 of the previous lemma can be summarized in the following.
\begin{thm}\label{eq_equiv}
For every type $A$, $\coeqrel_A$ is an equivalence relation on $\copar{A}$.
\end{thm}

It follows that the pair $\langle \copar{A}, \coeqrel_A \rangle$ is a total setoid \cite{BCP:2003}.

The {\em finite\/} elements of $\copar{A}$ are those that have a value in $A$.
They can be characterized directly by an inductive predicate.
\begin{defi}
$$
\begin{inductive}{\finitesym}{[x\colon \copar{A}]\colon \prop}
\finitereturn\colon (a\colon A)(\finitesym\ \return{a})\\
\finitestep  \colon (x\colon \copar{A})(\finitesym\ x)
                    \rightarrow (\finitesym\ \step{x}).
\end{inductive}
$$
\end{defi}
\begin{lem}
$
\forall x\colon \copar{A}.
(\finitesym\ x) \leftrightarrow 
\exists a\colon A.\parvalue{x}{a}.
$
\end{lem}
\proof 
From left to right: by induction on the proof of $(\finitesym\ x)$.
From right to left: by induction on the proof of $\parvalue{x}{a}$.
\qed

Hereafter, we use the notation $\finite{x}$ for $(\finitesym\ x)$.

A partial function from a type $A$ to a type $B$ is seen as a function from $\copar{A}$ to $\copar{B}$.
To be precise, we must require that a function $f\colon \copar{A}\rightarrow \copar{B}$ preserves equality:
$
\forall x_1,x_2\colon \copar{A}.
x_1\coeq x_2 \rightarrow (f\ x_1) \coeq (f\ x_2).
$
In other words, $f$ must be a setoid function.
Using the notation of \cite{capretta:1999}, this is written
$$
\langle \copar{A},\coeqrel_A \rangle
\mathrel{[{\rightarrow}]}
\langle \copar{B},\coeqrel_B \rangle.
$$
($[{\rightarrow}]$ is the {\em functionoid\/} constructors: Given two setoids, it constructs the setoid of functions between the two.
Its elements are pairs consisting of a function and a proof that it preserves the setoid equality.
The equality of the functionoid is the extensional equality on the first component, that is, the underlying function.)
Anyway, we will use the simple definition and leave it to the reader to check that all the functions that we define (with one exception) preserve equality.

A function is called {\em strict\/} if it always maps diverging elements to diverging elements.
Since a strict function is always determined on the diverging elements and it must preserve equality, its type can be strengthened to $A\rightarrow \copar{B}$.

The first partial functions that we define are just the lifting of the total functions.
\begin{defi}\label{par_lift}
If $f\colon A\rightarrow B$, we define its {\em lifting\/} to the partial elements as
$$
\begin{array}{l}
\parlift{f}\colon \copar{A} \rightarrow \copar{B}\\
\parlift{f}\ \return{a} = \return{(f\ a)}\\
\parlift{f}\ \step{y}   = \step{(\parlift{f}\ y)}
\end{array}
$$
\end{defi}
\begin{lem}
For every $a\colon A$, $(\parlift{f}\ \return{a}) \coeq \return{(f\ a)}$.
\end{lem}

A special case arises when we consider tuples and projections.
The types $\copar{A}\times\copar{B}$ and $\copar{(A\times B)}$ are not isomorphic.
We want to define a strict version of the projection functions.
That is, the projection should diverge whenever either of the components of the tuple diverges.
To do this, we first map $\copar{A}\times\copar{B}$ to $\copar{(A\times B)}$:
$$
\begin{array}{l}
\mixedpair\colon A\times \copar{B} \rightarrow \copar{(A\times B)}\\
\mixedpair\ \langle a,y\rangle =
  \begin{bycases}{y}
  \return{b} & \return{\langle a,b\rangle}\\
  \step{y'}  & \step{(\mixedpair\ \langle a,y'\rangle)}
  \end{bycases}\\
\ \\
\strictpair\colon \copar{A}\times\copar{B} \rightarrow \copar{(A\times B)}\\
\strictpair\ \langle x,y\rangle =
  \begin{bycases}{x}
  \return{a} & (\mixedpair\ \langle a,y\rangle)\\
  \step{x'}  & \step{(\strictpair\ \langle x',y\rangle)}
  \end{bycases}
\end{array}
$$
The function $\strictpair$ just moves all the steps outside the pairing constructor.
Iterating $\strictpair$ $n$ times, we obtain $\strictuple_n\colon (\copar{A})^n \rightarrow \copar{(A^n)}$.
We use the notation $\stuple{x_1,\ldots,x_n}$ for $(\strictuple_n\ \langle x_1,\ldots,x_n\rangle).$
The {\em strict projections\/} are then defined by
$$
\begin{array}{l}
\pistrict{i}\colon (\copar{A})^n \rightarrow \copar{A}\\
\pistrict{i} = \parlift{(\pi_i)} \circ \strictuple_n
\end{array}
$$
where $\pi_i\colon A^n \rightarrow A$ is the standard projection.
Note that $(\pistrict{i}\ \stuple{x_1,\ldots,x_n}) = x_i$ if and only if all the $x_j$s converge or $x_i$ diverges.

\section{Representation of partial recursive functions}\label{sec:recursion}
We show that every partial recursive function $f\colon \nat^n\pararrow \nat$ can be implemented in type theory as $\recfun{f}\colon (\conat)^n\rightarrow \conat$.
We use the notation $A \pararrow B$ to denote partial recursive functions from $A$ to $B$, that is, any function obtained from the base functions zero, successor, and projections, by composition, primitive recursion, and minimization.
Let us define $\recfun{f}$ for every basic function and function-forming operations (see \cite{phillips:1992} or any other generic introduction to recursion theory).
\begin{defi}\label{def:recfunctions}
By recursion on the definition of a partial recursive function $f\colon \nat^n\pararrow \nat$, we define its type-theoretic version $\recfun{f}\colon (\conat)^n\rightarrow \conat$.
\begin{description}
\item[Zero]
The zero function is the constant $0$:
$$
\begin{array}{l}
\underline{0}\colon \nat \rightarrow \nat\\
\underline{0}\ x = 0.
\end{array}
$$
Since $\underline{0}$ is a total function definable in type theory, we just put $\recfun{\underline{0}} = \parlift{\underline{0}}$.
\item[Successor]
Since the successor function $\s$ is definable in type theory, we just put $\recfun{\s} = \parlift{\s}$.
\item[Projections]
The projections are the functions $\pistrict{i}$ defined in the previous section.
\item[Composition]
Composition is the standard composition in type theory.
If $f\colon \nat^k\pararrow\nat$ and $g_i\colon \nat^n\pararrow \nat$ for $1\leq i\leq k$, then we define
$$
\begin{array}{l}
\recfun{(f\circ \langle g_1, \ldots, g_k \rangle)}
                \colon (\conat)^n\rightarrow \conat\\
\recfun{(f\circ \langle g_1, \ldots, g_k \rangle)}\ \seq{x} =
(\recfun{f}\ \langle (\recfun{g_1}\ \seq{x}), \ldots, (\recfun{g_k}\ \seq{x})\rangle).
\end{array}
$$
\item[Primitive Recursion]
Let $f\colon \nat^n\pararrow \nat$ and $g\colon \nat^{n+2}\pararrow \nat$, and let $h$ be defined by primitive recursion from $f$ and $g$:
$$
\begin{array}{l}
h\colon \nat^{n+1}\pararrow \nat\\
h\ \langle \seq{x}, 0\rangle = (f\ \seq{x})\\
h\ \langle \seq{x}, (\s\ y)\rangle =
    (g\ \seq{x}\ y\ (h\ \seq{x}\ y)).
\end{array}
$$
We first define a version $h'$ of $h$ by recursion (fixpoint) on the natural numbers:
$$
\begin{array}{l}
h'\colon (\conat)^n\rightarrow \nat\rightarrow \conat\\
h'\ \seq{x}\ 0 = (\recfun{f}\ \seq{x})\\
h'\ \seq{x}\ (\s\ m) = (\recfun{g}\ \seq{x}\ \return{m}\ (h'\ \seq{x}\ m))
\end{array}
$$
and then lift it to the partial elements:
$$
\begin{array}{l}
\recfun{h}\colon (\conat)^{n+1}\rightarrow \conat\\
\recfun{h}\ \seq{x}\ \return{m} = (h'\ \seq{x}\ m)\\
\recfun{h}\ \seq{x}\ \step{y}   = \step{(\recfun{h}\ \seq{x}\ y)}.
\end{array}
$$
\item[Minimization]
Let $f\colon \nat^{n+1}\pararrow \nat$ and let $g$ be defined from $f$ by minimization:
$$
\begin{array}{l}
g\colon \nat^n\pararrow \nat\\
g\ \seq{x} = \mbox{least $y$ such that }(f\ \seq{x}\ y)=0.
\end{array}
$$
To define the type-theoretic version of $g$ we use an auxiliary function that has an extra accumulation parameter, defined by corecursion (cofixpoint) on the result:
$$
\begin{array}{l}
g'\colon (\conat)^n\rightarrow \nat\rightarrow\conat\rightarrow\conat\\
g'\ \seq{x}\ i\ \return{m} =
  \begin{bycases}{m}
  0 & \return{i}\\
  (\s\ m') & \step{(g'\ \seq{x}\ (\s\ i)\ (\recfun{f}\ \langle \seq{x},\return{(\s\ i)}\rangle))}
  \end{bycases}\\
g'\ \seq{x}\ i\ \step{y} = \step{(g'\ i\ y)}
\end{array}
$$
and then
$$
\begin{array}{l}
\recfun{g}\colon (\conat)^n \rightarrow \conat\\
\recfun{g}\ \seq{x} = (g'\ \seq{x}\ 0\ (\recfun{f}\ \langle \seq{x},\return{0}\rangle)).
\end{array}
$$
\end{description}
\end{defi}

It is routine work, although quite long, to verify that the translations have the correct computational behavior.
\begin{thm}\label{th:recursion}
Let $f\colon \nat^n\pararrow\nat$ be a partial recursive function.
For every $\seq{x}\colon \nat^n$ and $y\colon \nat$,
$$
(f\ \seq{x})=y
\quad \Longleftrightarrow \quad
(\recfun{f}\ \seq{\return{x}}) \coeq \return{y}.
$$
\end{thm}
\proof 
The proof is a lengthy routine use of the techniques illustrated in the previous section.
For a formal proof we should first formalize part of recursion theory.
First, we have to define an inductive type of codes of recursive functions generated by constructors corresponding to the base functions, composition, primitive recursion, and minimization.
Then, we need to give an operational semantics that associates to every code a relation on natural numbers, the relation being the graph of the recursive function.
Finally, we have to interpret the codes as functions on the type of partial elements, as shown above, and prove that this interpretation is sound with respect to the operational semantics.

There are no conceptual problems in doing this, but a lot of technical work.
We rather refer to Section \ref{sec:fixpoints}, where we prove that it is possible to construct fixed points of functional operators.
Since it is a known fact that all recursive functions can be realized by such a fixpoint combinator, the present statement will automatically follow.  
\qed

Let me stress the advantage of this approach in comparison with other methods to formalize general recursion in type theory.

First of all, some of the techniques, for example that of Balaa and Bertot \cite{balaa/bertot:2000} and that of Barthe and others \cite{BFGPU:2002,bgp:2005}, do not address the question of partiality but present ways of extending the definition schemes for total recursive functions.
The method of Bove and Capretta \cite{bove:2001,bove/capretta:2001,bove/capretta:2004a,bove/capretta:2005a} allows the definition of partial functions by restricting them to their domain of convergence.
However, it is still not possible to apply a function freely to an argument, but it is necessary first to prove that the argument satisfies the domain predicate.

With coinductively defined types of partial elements, functions can be freely applied to arguments without the need of extra logical information.

\section{Nested recursion}\label{sec:nested}
Theorem \ref{th:recursion} says that every computable function can be represented in type theory.
In this section we look at some specific examples of nested recursive functions.
In general we represent a partial function from $A$ to $B$ as an element of $A\rightarrow \copar{B}$, since the values on the diverging elements of $A$ must, by strictness, diverge.

We start with the simplest example of nested recursion \cite{bove/capretta:2001}:
$$
\begin{array}{l}
\ttf{nest}\colon \nat \pararrow \nat\\
\ttf{nest}\ 0 = 0\\
\ttf{nest}\ (\s\ n) = (\ttf{nest}\ (\ttf{nest}\ n)).
\end{array}
$$
It is clear that $\ttf{nest}$ is constantly $0$, so we can implement it in type theory as the constant $0$.
However, we are interested in the form in which the function is defined.
More complex nested functions do not have a simple non-nested presentation.
We see an example later.
We represent $\ttf{nest}$ in type theory by using the method of accumulation of results and tail recursion:
$$
\begin{array}{l}
\ttf{cnest}\colon \nat\times \nat \rightarrow \conat\\
\ttf{cnest}\ \langle n,0\rangle = \return{n}\\
\ttf{cnest}\ \langle 0, (\s\ m)\rangle =
    \step (\ttf{cnest}\ \langle 0,m \rangle)\\
\ttf{cnest}\ \langle (\s\ n),(\s\ m)\rangle =
    \step (\ttf{cnest}\ \langle n,m+2 \rangle).
\end{array}
$$
It is easy to check that $(\ttf{cnest}\ \langle n,m\rangle)$ computes the value $(\ttf{nest}^m\ n)$, so the extra parameter $m$ keeps track of the number of nested iteration of $\ttf{nest}$.
Now we can define in type theory
$$
\begin{array}{l}
\ttf{nest}\colon \nat\rightarrow \conat\\
\ttf{nest}\ n = (\ttf{cnest}\ \langle n,1 \rangle).
\end{array}
$$

Let us look at a more interesting example.
We define a class of nested recursive functions, that we nickname {\em the devil's nest\/}.
Let $A$ be a type, $T$ a decidable subset of $A$, representable in type theory by a predicate $P\colon A\rightarrow \prop$ such that $\forall a:A. (P\ a)\lor \neg (P\ a)$.
Let $i\colon A\rightarrow A$ and $g\colon T \rightarrow A$ (that is, $g\colon (x\colon A)(P\ x) \rightarrow A$).
Then the function $\ttf{dev}_{T,i,g}$ is defined as
$$
\begin{array}{l}
\ttf{dev}_{T,i,g}\colon A \pararrow A\\
\ttf{dev}_{T,i,g}\ a = \begin{cases}
       (g\ a) & \mbox{if }a\in T\\
       (\ttf{dev}_{T,i,g}\ (\ttf{dev}_{T,i,g}\ (i\ a)) & \mbox{otherwise}
       \end{cases}
\end{array}
$$
We formalize this function in type theory by using, as before, an extra parameter that keeps track of the number of nested calls to $\ttf{dev}_{T,i,g}$:
$$
\begin{array}{l}
\ttf{dev_{aux}}\colon A\times\nat \rightarrow \copar{A}\\
\ttf{dev_{aux}}\ \langle a,m \rangle\\ =
  \begin{cases}
  \left(\begin{bycases}{m}
        0 & \return{(g\ a)}\\
        (\s\ m') & \step{(\ttf{dev_{aux}}\ \langle (g\ a),m' \rangle)}
        \end{bycases}
  \right) & \mbox{if }(P\ a)\\
  \step{(\ttf{dev_{aux}}\ \langle (i\ a),(\s\ m)\rangle)} & \mbox{otherwise}
  \end{cases}
\end{array}
$$
We have left the proof argument of $g$ implicit to simplify notation: Rigorously, we should have written $(g\ a\ h)$ in place of $(g\ a)$, in the case where $(P\ a)$ is true, where $h$ is the proof of $(P\ a)$ guaranteed by the branch case.
Similarly to the previous case, the original function is recovered by specifying an initial value for the accumulation parameter, in this case $0:$
$$
\begin{array}{l}
\ttf{dev}_{T,i,g}\colon A \rightarrow \copar{A}\\
\ttf{dev}_{T,i,g}\ a = (\ttf{dev_{aux}}\ \langle a,0\rangle).
\end{array}
$$

Another interesting example occurs when a recursive call is followed by a call to another function, as in the definition of primitive recursion but without a decreasing argument.
Let $h\colon A\rightarrow A$.
We want to formalize the following function:
$$
\begin{array}{l}
\ttf{d}_h\ a =
\begin{cases}
(g\ a) & \mbox{if }x\in T\\
(h\ (\ttf{d}_h\ (i\ a))) & \mbox{otherwise}.
\end{cases}
\end{array}
$$
We use a continuation-passing style translation, using an extra functional parameter:
$$
\begin{array}{l}
\ttf{d_{aux}}\colon (A\rightarrow \copar{A})\times A \rightarrow \copar{A}\\
\ttf{d_{aux}}\ \langle k,a \rangle =
\begin{cases}
k\ (g\ a) & \mbox{if }(P\ a)\\
\step{(\ttf{d_{aux}}\ \langle k\circ h, (i\ a)\rangle)} & \mbox{otherwise}
\end{cases}
\end{array}
$$
and then
$$
\begin{array}{l}
\ttf{d}_h\colon A\rightarrow \copar{A}\\
\ttf{d}_h\ a = (\ttf{d_{aux}}\ \langle \lambda x.\return{x},a \rangle).
\end{array}
$$

Finally, we put the two preceding examples together to obtain the most general version of the devil's nest:
$$
\begin{array}{l}
\ttf{devil}_{T,i,g,h}\colon A \pararrow A\\
\ttf{devil}_{T,i,g,h}\ a = \begin{cases}
       (g\ a) & \mbox{if }a\in T\\
       (h\ (\ttf{devil}_{T,i,g,h}\ (\ttf{devil}_{T,i,g,h}\ (i\ a))) & \mbox{otherwise}
       \end{cases}
\end{array}
$$
This function is formalized by
$$
\begin{array}{l}
\ttf{devil_{aux}}\colon (A\rightarrow\copar{A})\times \nat\times A
         \rightarrow \copar{A}\\
\ttf{devil_{aux}}\ \langle k,m,a \rangle \\ =
\begin{cases}
\left(\begin{bycases}{m}
      0 & k\ (g\ a)\\
      (\s\ m') & \step{(\ttf{devil_{aux}}\ \langle k, m', (g\ a) \rangle)}
      \end{bycases}
\right)
  & \mbox{if }(P\ a)\\
\step{(\ttf{devil_{aux}}\ \langle k\circ h, (\s\ m), (i\ a) \rangle)}
  & \mbox{otherwise}
\end{cases}
\end{array}
$$
and then
$$
\begin{array}{l}
\ttf{devil}_{T,i,g,h}\colon A \rightarrow \copar{B}\\
\ttf{devil}_{T,i,g,h}\ a = (\ttf{devil_{aux}}\ \langle \lambda x.\return{x},0,a \rangle).
\end{array}
$$

There is another possible generalization: Instead of having just one nested recursive call, we could have a variable number of nestings.
It is clear that then we just have to modify the definition by adding the nesting number to the accumulation parameter.

\section{Fixed points of function operators}\label{sec:fixpoints}

We want to construct functions defined as fixed points of function operators.
Let $F\colon (A\rightarrow \copar{B})\rightarrow (A\rightarrow \copar{B})$ be such an operator.
Our goal is to define a function $\Y(F)\colon A\rightarrow \copar{B}$ such that $(F\ \Y(F))$ is extensionally equal to $\Y(F)$.
Moreover $\Y(F)$ must be minimal with respect to convergence.

Some conditions must be imposed on $F$ for the construction to be possible.
Since $F$ is supposed to represent a generic recursive scheme, one sensible condition is finitarity: We assume that, to compute a specific result, $F$ uses its arguments only on a finite number of inputs.
\begin{defi}\label{def:finitary}
We say that $F\colon (A\rightarrow \copar{B})\rightarrow (A\rightarrow \copar{B})$ is {\em finitary\/} if it satisfy the following condition:
For every function $f\colon A\rightarrow \copar{B}$ and every argument $a\colon A$ such that $(F\ f\ a)$ converges to some value, that is,  $\parvalue{(F\ f\ a)}{b}$ for some $b\colon B$; there exists a finite number of arguments $a_1, \ldots, a_n\colon A$ such that $f$ converges on each of them, $\parvalue{(f\ a_i)}{b_i}$, and
$$
\begin{array}{l}
\forall g\colon A\rightarrow \copar{B}.
\bigwedge_{i=1}^{n} \parvalue{(g\ a_i)}{b_i} \rightarrow
\parvalue{(F\ g\ a)}{b}
\end{array}
$$
In words, $F$ is finitary (or continuous) if its results depend only on the values of the function argument on a finite set of inputs.
\end{defi}
Notice that all operators used in recursion theory, for example those needed for the constructions of Section \ref{sec:recursion}, are finitary.

We will use three consequences of finitarity: first, a finitary operator preserves extensional equality; second, it preserves convergence order; third, its least prefixed point can be constructed in a countable number of steps.

In the rest of this section, let $F$ be a finitary operator.

\begin{defi}
{\em Extensional equality\/} between two functions $f_1, f_2\colon A\rightarrow \copar{B}$ is defined as
$$
f_1 \exteq f_2 \Longleftrightarrow \forall a\colon A. (f_1\ a) \coeq (f_2\ a).
$$

We say that $F$ is {\em extensional\/} if
$$
\forall f_1, f_2\colon A\rightarrow \copar{B}. f_1\exteq f_2 \rightarrow (F\ f_1)\exteq (F\ f_2).
$$
\end{defi}

\begin{lem}\label{lemma:finitary_extensional}
$F$ is extensional.
\end{lem}
\proof 
It follows trivially from the stronger Lemma \ref{lemma:F_order}.
\qed

A stronger property holds: $F$ preserves the order on functions given by convergence.

\begin{defi}
The {\em convergence order\/} between partial elements is defined coinductively by
$$
\begin{bigcoinductive}{\colerel_A}{[x,y\colon \copar{A}]\colon \prop}
\colevalue\colon (x,y\colon \copar{A}; a\colon A)
                 \parvalue{x}{a} \rightarrow \parvalue{y}{a}
                 \rightarrow (\colerel_A\ x\ y)\\
\colesteps\colon (x,y\colon \copar{A})
                 (\colerel_A\ x\ y) \rightarrow
                 (\colerel_A\ \step{x}\ \step{y})\\
\colelstep\colon (x,y\colon \copar{A})
                 (\colerel_A\ x\ y) \rightarrow
                 (\colerel_A\ \step{x}\ y)
\end{bigcoinductive}
$$
We use the notation $x\cole y$ for $\colerel_A\ x\ y.$

Intuitively, $x\cole y$ holds if $x$ is obtained from $y$ by adding some (potentially infinite) $\step$ steps.

The order between functions is defined pointwise on their values, that is, if $f_1, f_2\colon A\rightarrow \copar{B}$, then we define
$$
f_1 \cole f_2 \Longleftrightarrow \forall a\colon A, (f_1\ a) \cole (f_2\ a).
$$
\end{defi}

It is immediate that equality ${}\coeq{}$ is a subrelation of the order ${}\cole{},$ since the order is defined by three constructors of which the first two correspond to the constructors of equality.
It is also easy to prove that ${}\cole{}$ is a transitive relation and it is reflexive and antisymmetric with respect to ${}\coeq{}.$
The relation ${}\cole{}$ is equivalent to implication between convergence statements.
\begin{lem}\label{lemma:cole_imp}
The proposition $x\cole y$ is equivalent to $\forall b\colon B. \parvalue{x}{b} \rightarrow \parvalue{y}{b}.$ 
\end{lem}
\proof 
From left to right, by induction on the proof of $\parvalue{x}{b}$.
From right to left, by coinduction and cases on $x$.
\qed

By antisymmetry we can conclude that the following characterization of equality holds.
\begin{lem}\label{lemma:eq_convergence}
The proposition $x\coeq y$ is equivalent to $\forall  b\colon B. \parvalue{x}{b} \leftrightarrow \parvalue{y}{b}.$ 
\end{lem}

\begin{lem}\label{lemma:F_order}
The operator $F$ preserves the convergence order, that is,
$$
\forall f_1, f_2\colon A\rightarrow \copar{B}.
f_1\cole f_2 \rightarrow (F\ f_1)\cole (F\ f_2).
$$
\end{lem}
\proof 
It is a straightforward consequence of finitarity.
\qed

We want to define the least fixed point $\Y(F)\colon A \rightarrow \copar{B}$.
Intuitively, we run in parallel all the iterations of $F$, starting with the always undefined function, and we take as result the outcome of the first converging run.
If we call $\bot$ the function $\lambda a. \stepfun^\infty$, then we have:
$$
\Y(F)\ a =
\begin{array}[t]{ll}
(\bot\ a) & \mbox{if it converges; otherwise}\\
((F\ \bot)\ a) & \mbox{if it converges; otherwise}\\
((F^2\ \bot)\ a) & \mbox{if it converges; otherwise}\\
\quad \vdots
\end{array}
$$
(This intuitive explanation is not precise: Since we cannot decide convergence, the actual function choses the first of those values that converges.)
Formally, we start by defining a function that computes the first converging of two partial objects.
\comment{
 we can define it as
$$
\begin{array}{l}
f'\colon (A\rightarrow \copar{B}) \times \copar{B}\times A
         \rightarrow \copar{B}\\
f'\ \langle k,\return{b},a \rangle = \return{b}\\
f'\ \langle k,\step{x},a \rangle
  = \step{(f'\ \langle (F\ k), (\ttf{fstconv}\ x\ (k\ a)),a \rangle)}
\end{array}
$$
where
}

\begin{defi}
The function computing the first converging element of a pair is defined corecursively by
$$
\begin{array}{l}
\ttf{fstconv}\colon \copar{B}\rightarrow\copar{B}\rightarrow\copar{B}\\
\ttf{fstconv}\ \return{b}\ y = \return{b}\\
\ttf{fstconv}\ \step{x}\ \return{b} = \return{b}\\
\ttf{fstconv}\ \step{x}\ \step{y} = \step{(\ttf{fstconv}\ x \ y)}.
\end{array}
$$
We use the notation $\fstconv{x}{y}$ for $(\ttf{fstconv}\ x\ y).$
\end{defi}

So $\fstconv{x}{y}$ returns the first between $x$ and $y$ to converge.
Note that this is the only function defined so far that does not preserve equality: The term $\fstconv{(\stepfun^n{\return{b_1}})}{(\stepfun^m{\return{b_2}})}$ converges to $b_1$ if $n\leq m$, to $b_2$ otherwise, without $b_1$ and $b_2$ being necessarily equal.
The result of the function depends sensibly on the number of $\stepfun$ steps in the arguments.
However, our use of $\ttf{fstconv}$ to obtain fixed points is such that the arguments are always {\em compatible\/}, that is, they never converge to different elements, although it may be possible that one of the two converges while the other diverges.
In this case the result is equality preserving.
(The notion of compatibility of partial elements is due to Tarmo Uustalu.)

\comment{
Then we can define
$$
\begin{array}{l}
\Y(F)\colon A\rightarrow \copar{B}\\
\Y(F)\ a = (f'\ \langle \lambda x.\stepfun^\infty,\stepfun^\infty,a\rangle).
\end{array}
$$
}

\begin{lem}\label{lemma:fstconv}
For every $y\colon\copar{B}$,
$
(\fstconv{\stepfun^\infty}{y}) \coeq y.
$
\end{lem}
\proof 
Proof by coinduction. Assume $H$ is the proof of the statement:
$$
H\colon \forall y\colon \copar{B}.(\fstconv{\stepfun^\infty}{y}) \coeq y.
$$
In the computation of $\fstconv{\stepfun^\infty}{y}$ the first equation in the definition of $\ttf{fstconv}$ is not used.
The second or third equation is applied, according to the structure of $y$.
If $y=\return{b}$, then the second equation itself states the truth of the lemma, by reflexivity of ${}\coeq{}.$
If $y=\step{y'}$ then, by the third equation, $(\fstconv{\stepfun^\infty}{y}) = (\fstconv{\step{\stepfun^\infty}}{\step{y'}}) =  \step{(\fstconv{\stepfun^\infty}{y'})}$.
By the coinductive hypothesis $H$ applied to $y'$, $\fstconv{\stepfun^\infty}{y'} \coeq y'$.
By $\coeqstep$ we then have that $\step{(\fstconv{\stepfun^\infty}{y'})} \coeq \step{y'}$, that is, $(\fstconv{\stepfun^\infty}{y}) \coeq y$, as desired.
The recursive application of $H$ is guarded by $\coeqstep$.
\qed

Obviously, $\ttf{fstconv}$ converges only if one of its arguments converges.
\begin{lem}
For every $x, y\colon\copar{B}$ and $b\colon B$, 
$
\parvalue{(\fstconv{x}{y})}{b} \rightarrow
  \parvalue{x}{b}\lor \parvalue{y}{b}.
$
\end{lem}
\proof 
By induction on the proof of $\parvalue{(\fstconv{x}{y})}{b}.$
\qed

The vice-versa is also true, but we have to be careful to take into account the non-extensionality of $\ttf{fstconv}$: if, for example, $x$ converges to $b$, it is not guaranteed that $\fstconv{x}{y}$ also converges to $b$, because $y$ may converge to a different $b'$ in a shorter time.
However, if $x$ is lower that $y$ in the convergence order, we know that they cannot converge to different values.
\begin{lem}
Let $x, y\colon \copar{B}$ and $b\colon B$, assume that $x\cole y$; then
$\parvalue{x}{b} \rightarrow \parvalue{(\fstconv{x}{y})}{b}.$
\end{lem}
\proof 
By induction of the proof of $\parvalue{x}{b}.$
\qed

We can recursively define an infinitary version of $\ttf{fstconv}$.
\begin{defi}
The operation of computing the first converging element of a sequence is defined corecursively by
$$
\begin{array}{l}
\ttf{parallel\_search\_aux}\colon
  (\nat\rightarrow \copar{B}) \rightarrow \nat \rightarrow \copar{B}
  \rightarrow \copar{B}\\
\ttf{parallel\_search\_aux}\ f\ n\ \return{b} = \return{b}\\
\ttf{parallel\_search\_aux}\ f\ n\ \step{x} = \step{\ttf{parallel\_search\_aux}\ f\ (\s\ n)\ (\fstconv{x}{(f\ n)})}\\
\ \\
\ttf{parallel\_search}\colon (\nat\rightarrow\copar{B})\rightarrow\copar{B}\\
\ttf{parallel\_search}\ f = \ttf{parallel\_search\_aux}\ f\ 0\ \stepfun^\infty.
\end{array}
$$
We write $(\infconvaux{f}{n}{x})$ for $(\ttf{parallel\_search\_aux}\ f\ n\ x)$ and $\infconv{f}$ for $(\ttf{parallel\_search}\ f)$.
\end{defi}

Infinitary versions of the lemmas that we proved for $\ttf{fstconv}$ hold for $\ttf{parallel\_search}$, with corresponding additional hypotheses to take non-extensionality into account.
\begin{lem}\label{lemma:parallel_search_converge}
For all $f\colon \nat\rightarrow\copar{B}$ and $b\colon B$;
$\parvalue{(\infconv{f})}{b} \rightarrow \exists n. \parvalue{(f\ n)}{b}.$
\end{lem}
\begin{lem}\label{lemma:parallel_search_converge_n}
Let $f\colon\nat\rightarrow\copar{B}$ be increasing, that is, $\forall n, m. n<m \rightarrow (f\ n)\cole (f\ m)$; if $b\colon B$ and $n\colon \nat$, then
$\parvalue{(f\ n)}{b} \rightarrow \parvalue{(\infconv{f})}{b}.$
\end{lem}
\begin{lem}\label{lemma:parallel_search_bound}
Let $f\colon \nat \rightarrow \copar{B}$ and $y\colon \copar{B}$;
if $\forall n\colon \nat. (f\ n)\cole y$, then  $\infconv{f} \cole y.$
\end{lem}

Let us call $k_i$ the $i$th iteration of $F$ on the function that always diverges:
$$
\begin{array}{l}
k_0 \equiv \lambda x.\stepfun^\infty\\
k_{n+1} \equiv (F\ k_n).
\end{array}
$$

Convergence of the $k_i$s is stable with respect to the index $i$.
\begin{lem}\label{lemma:ks}
For every $n, m\colon\nat$ such that $n\leq m$, $k_n \cole k_{m}$;
equivalently
$$
\forall a\colon A, b\colon B. \parvalue{(k_n\ a)}{b} \Rightarrow \parvalue{(k_m\ a)}{b}.
$$ 
\end{lem}
\proof 
We prove by induction on $n$ that $k_n\cole k_{n+1}$, the statement follows by transitivity of ${}\cole{}.$
The base case is obvious since $k_0$ has the constant value $\step^\infty$, the least element of the convergence order.
The inductive step follows immediately from Lemma \ref{lemma:F_order}.
\qed

Finally, we can simply define the fixed point of $F$ pointwise as the result of running all of the $k_i$s in parallel.
\begin{defi}
$
\Y(F) = \lambda a\colon A. \infconv{(\lambda n. (k_n\ a))}.
$
\end{defi}

If $\Y(F)$ converges on a certain element $a$, it must give the same result as one of the $k_i$s.
\begin{lem}\label{lemma:yk1}
$
\parvalue{(\Y(F)\ a)}{b} \Leftrightarrow \exists n.\parvalue{(k_n\ a)}{b}.
$
\end{lem}
\proof 
From left to right it follows from Lemma \ref{lemma:parallel_search_converge}.
From right to left it follows from Lemma \ref{lemma:parallel_search_converge_n}.
\qed

The combination of the previous lemmas provides a proof of the soundness of $\Y(F)$ as a least fixed point of $F$.
\begin{thm}\label{thm:fixed_point}
$\Y(F)$ is a fixed point of $F$: $(F\ \Y(F)) \exteq \Y(F)$.
\end{thm}
\proof 
We prove that $(F\ \Y(F)) \cole \Y(F)$ and $\Y(F) \cole (F\ \Y(F))$.
The statement follows from antisymmetry of ${}\cole{}$.

To prove that $(F\ \Y(F)) \cole \Y(F)$ we just have to show, by Lemma \ref{lemma:cole_imp}, that for all $a\colon A$ and $b\colon B$, if $\parvalue{(F\ \Y(F)\ a)}{b}$ then also $\parvalue{(\Y(F)\ a)}{b}.$
By finitarity of $F$, $\parvalue{(F\ \Y(F)\ a)}{b}$ implies that there exist $a_1,\ldots, a_k\colon A$ and $b_1,\ldots,b_k\colon B$ such that $\parvalue{(\Y(F)\ a_i)}{b_i}$ and the result of $(F\ \Y(F)\ a)$ depends only on these arguments.
By Lemma \ref{lemma:yk1}, we have that $\parvalue{(k_{n_i}\ a_i)}{b_i}$ for some indexes $n_i.$
Let $n$ be the largest of the $n_i$s.
By Lemma \ref{lemma:ks}, $\parvalue{(k_n\ a_i)}{b_i}$ for every $i.$
Therefore, by finitarity of $F$, $\parvalue{(F\ k_n\ a)}{b}.$
But $(F\ k_n) = k_{n+1}$, so $\parvalue{(k_{n+1}\ a)}{b}$.
By Lemma \ref{lemma:yk1}, it follows that $\parvalue{(\Y(F)\ a)}{b}$, as desired.

In the other direction, we prove that $\Y(F) \cole (F\ \Y(F))$.
Let $a\colon A$, we show that $(k_n\ a)\cole (F\ \Y(F)\ a)$ for every $n$; the statement follows by Lemma \ref{lemma:parallel_search_bound}.
For $n=0$ it is trivial because $(k_0\ a)=\stepfun^\infty$.
For non-zero values we have $(k_{n+1}\ a) = (F\ k_n\ a)$ and, by Lemma \ref{lemma:F_order}, we just need to prove $(k_n\ a) \cole (\Y(F)\ a).$
But this follows easily from Lemmas \ref{lemma:cole_imp} and \ref{lemma:yk1}.
\qed

\begin{thm}\label{thm:least_prefix}
$\Y(F)$ is the least prefixed point of $F$:
if $f$ is any prefixed point of $F$, that is, $(F\ f)\cole f$, then $\Y(F)\cole f.$
\end{thm}
\proof 
We prove by induction on $n$ that $k_n\cole f$.
For $n=0$ it is obvious.
Assume that $k_n\cole f.$
Then, by Lemma \ref{lemma:F_order}, $k_{n+1} = (F\ k_n) \cole (F\ f) \cole f$ and we are done.
The statement then follows from Lemma \ref{lemma:parallel_search_bound}.
\qed

A method to construct fixed points of function operators is described by Balaa and Bertot \cite{balaa/bertot:2000}, but in their case it is necessary to prove that recursive calls are decreasing with respect to some wellfounded order.
The advantage of our method is that it produces fixed points of every finitary operator, without requiring any additional logical information.

\section{Lazy interpretation}\label{sec:lazy}
The formalization of partial elements of $A$ as terms of type $\copar{A}$ gives a {\em strict\/} interpretation of functions.
There cannot be a partial evaluation of a term: If $x\colon \copar{A}$, then we can investigate its shape, that is either a result $\return{a}$ or a step $\step{x'}$.
In the latter case, we can go on investigating the shape of $x'$.
As long as we get step cases, we have no information about the result.
When we get a result $\return{a}$ we get all the information.

In functional programming, it is useful to be able to compute only partial information about a result.
For example, we may need to know that a certain natural number result is a successor, without computing it completely.
An instance in which this capacity can be used in computation is the following example of two mutually defined recursive functions:
$$
\begin{array}{l}
\slothf\colon \nat\pararrow\nat\\
\slothf\ 0 = 0\\
\slothf\ (\s\ n) = (\slothf\ (\slothg\ n)) + (\slothg\ n)\\
\\
\slothg\colon \nat\pararrow\nat\\
\slothg\ 0 = 0\\
\slothg\ (\s\ m) =
\begin{cases}
(\slothg\ (\slothf\ m))+m & \mbox{if }(\slothf\ m)\leq m\\
0                         & \mbox{otherwise}
\end{cases}
\end{array}
$$
It is easy to see that the computation of $(\slothf\ 13)$ diverges, independently of whether the evaluation is strict or lazy.
However, if we evaluate it partially, we obtain
$$
(\slothf\ 13) = (\slothf\ (\slothg\ 12)) + (\slothg\ 12)
              = (\slothf\ 19) + 19.
$$
When computing $(\slothg\ 14)$, we need to decide whether $(\slothf\ 13)\leq 13$.
With a strict evaluation strategy, the computation of $(\slothg\ 14)$ diverges because the evaluation of $(\slothf\ 13)$ diverges.
However, with a lazy evaluation strategy, we can determine that $(\slothf\ 13) = (\slothf\ 19) + 19 > 13$ and therefore $(\slothg\ 14)=0$.
This shows that a lazy evaluation strategy may converge when a strict one diverges.

To formalize lazy evaluation we have to modify the formalization of partial elements for inductive types.
For an inductive type $\ttf{T}$, we define the type of its partial elements as the coinductive type with the same constructors plus a constructor for a computation step that does not yield any information.
\begin{defi}
Let $\ttf{T}$ be an inductive type, that is, $\ttf{T}$ is defined in type theory by
$$
\begin{inductive}{\ttf{T}}{\colon \type}
\ttf{c}_1 \colon (\Theta_1)\ttf{T}\\
\vdots \\
\ttf{c}_n \colon (\Theta_n)\ttf{T}.
\end{inductive}
$$
The type of {\em lazy partial elements\/} of $\ttf{T}$ is
$$
\begin{coinductive}{\lazypar{\ttf{T}}}{\colon \type}
\ttf{c}_1 \colon (\Theta_1[\ttf{T}:=\lazypar{\ttf{T}}])\lazypar{\ttf{T}}\\
\vdots \\
\ttf{c}_n \colon (\Theta_n[\ttf{T}:=\lazypar{\ttf{T}}])\lazypar{\ttf{T}}\\
\stepsym \colon \lazypar{\ttf{T}}\rightarrow \lazypar{\ttf{T}}.
\end{coinductive}
$$
As before, we use the notation $\step{x}$ for $(\stepsym\ x)$.
We make a slight abuse of notation by using the same constructor names for $\ttf{T}$ and $\lazypar{\ttf{T}}.$
\end{defi}

This was my original formalization of partial elements (see Chapter 7 of \cite{capretta:2002}).
The version $\copar{A}$ is a simplification suggested to me by Herman Geuvers and Peter Aczel.

As an example, let us see how this variant produces the right computation behavior for the $\slothg$ function.
First of all, the type of lazy natural numbers is
$$
\begin{coinductive}{\lazynat}{\colon \type}
0\colon \lazynat\\
\s\colon \lazynat\rightarrow \lazynat\\
\step{} \colon \lazynat\rightarrow \lazynat.
\end{coinductive}
$$
Here are the lazy versions of addition and order, addition is defined by corecursion on its result and order is defined as an inductive relation:
$$
\begin{array}[t]{l}
\lazyplus{}{}\colon \lazynat \rightarrow \lazynat\rightarrow \lazynat\\
\lazyplus{x}{0} = x\\
\lazyplus{x}{(\s\ y)} = \s\ (\lazyplus{x}{y})\\
\lazyplus{x}{\step{y}} = \step{(\lazyplus{x}{y})}
\end{array}
\qquad
\begin{biginductive}{\lazyle}{\colon \lazynat\rightarrow\lazynat\rightarrow\prop}
\ttf{le_0}\colon (y\colon \lazynat) 0\lazyle y\\
\ttf{le_{\s}}\colon (x, y\colon\lazynat) x\lazyle y \rightarrow (\s\ x)\lazyle (\s\ y)\\ 
\ttf{le_{\stepsym,l}}\colon (x, y\colon\lazynat) x\lazyle y \rightarrow (\step{x})\lazyle y\\ 
\ttf{le_{\stepsym,r}}\colon (x, y\colon\lazynat) x\lazyle y \rightarrow x\lazyle (\step{y})\\ 
\end{biginductive}
$$

This definition of order is not reflexive, because it is not possible to prove $\stepfun^\infty \lazyle \stepfun^\infty$, but we adopt it for our example for simplicity.
The reader should find it easy to modify it into a reflexive relation.

Returning to our example, the conditioned equation in the definition of $\slothg$ produces the the statement $ \lazyplus{(\slothf\ 19)}{19} \lazyle 13$.
Contrary to the strict case, we can now evaluate this statement to a truth value.
By the definition of $\lazyplus{}{}$ this becomes
$$
\s^{19}(\slothf\ 19) \lazyle \s^{13}0.
$$ 
By inversion, if this statement were provable, the only applicable constructor would be $\le_{\s}$, so $\s^{18}(\slothf\ 19) \lazyle \s^{12}0$ should also be provable.
Repeating this step 13 times, we get $\s^{6}(\slothf\ 19) \lazyle 0$.
But this statement does not match the conclusion of any constructor of $\lazyle$, thus it is not provable.
We conclude that our original statement was not provable.

\section{Partiality as a monad}\label{sec:monad}
The categorical notion of (strong) monad is a useful abstract description of computation.
Eugenio Moggi studied this relation in a series of works \cite{moggi:1989,moggi:1991,bhm:2002}.
We will show that the operator mapping a type $A$ to the type of partial elements $\copar{A}$ gives a computational monad in Moggi's sense.
We recall the notion of monad in extension form, or Kleisli triple.
The definition is taken from \cite{bhm:2002}, Definition 3, pg.~45.
\begin{defi}[Kleisli triple/monad in extension form]
A {\em Kleisli triple} over a category $\mathcal{C}$ is a triple $(T,\eta,\_^*)$, where $T\colon |\mathcal{C}| \rightarrow |\mathcal{C}|$, $\eta_A\colon A\rightarrow TA$ for $A\in |\mathcal{C}|$, $f^*\colon TA\rightarrow TB$ for $f\colon A\rightarrow TB$ and the following equations hold:
\begin{itemize}
\item $\eta_A^* = \ttf{id}_{TA}$;
\item $f^* \circ \eta_A = f$ for $f\colon A\rightarrow TB$;
\item $g^*\circ f^* = (g^*\circ f)^*$ for $f\colon A\rightarrow TB$ and $g\colon B\rightarrow TC$.
\end{itemize}
\end{defi}

The intuitive understanding of a Kleisli triple is that, for a type of values $A$, $TA$ is the type of computations of elements of type $A$.
The unit $\eta_A$ maps a value $a$ to the trivial computation that just returns $a$.
A function $f\colon A\rightarrow TB$ that maps values of type $A$ to computations of type $B$ can be extended to computations: $f^*\colon TA\rightarrow TB$ is the program that, given a computation $x$ of type $A$, first computes it and, if it gives a value $a$, applies $f$ to it.

In the present case, $\mathcal{C}$ is the category of setoids in type theory \cite{BCP:2003}.
The operator $T$ maps a type $A$ to the type of partial elements $\copar{A}$.
To be precise, we must define this operation on setoids.
This is not difficult, it is just necessary to identify elements of $\copar{A}$ that give equal results according to the book equality of the setoid.
The details of the definition are left to the reader.
Also for the other components of the monad, we define them on types and leave to the reader the routine extension to setoids.
The unit of the monad $\eta$ is simply the $\returnsym$ constructor of $\copar{A}$.
If $f\colon A\rightarrow \copar{B}$, then $f^*\colon \copar{A}\rightarrow \copar{B}$ is defined by
$$
(f^*\ x) = \begin{bycases}{x}
           \return{a} & \return{(f\ a)}\\
           \step{x'}  & \step{(f^*\ x')}.
           \end{bycases}
$$
The equations for a Kleisli triple are satisfied. The proof of this fact is routine.
\begin{thm}
The triple $(\copar{\_},\returnsym,\_^*)$ is a Kleisli triple on the category of setoids.
\end{thm}

A more powerful notion is that of {\em strong monad\/}.
This is a monad in which a pair value-computation can be turned into the computation of a pair. The definition is taken from \cite{moggi:1989}.
\begin{defi}
A {\em strong monad\/} over a category $\mathcal{C}$ with finite products is a monad $(T,\eta,\mu)$ together with a natural transformation $t_{A,B}\colon A\times TB \rightarrow T(A\times B)$ satisfying the following equations:
\begin{itemize}
\item $T(r_A)\circ t_{1,A} = r_{TA}$;
\item $T(\alpha_{A,B,C})\circ t_{A\times B,C} = t_{A,B\times C}\circ (\ttf{id}_A\times t_{B,C}) \circ \alpha_{A,B,TC}$;
\item $t_{A,B}\circ (\ttf{id}_A\times \eta_B) = \eta_{A\times B}$;
\item $t_{A,B}\circ (\ttf{id}_A\times \mu_B) = \mu_{A\times B}\circ T(t_{A,B}) \circ t_{A,TB}$.
\end{itemize}
where $r$ and $\alpha$ are the natural isomorphisms
\begin{itemize}
\item $r_A\colon 1\times A\rightarrow A$;
\item $\alpha_{A,B,C}\colon (A\times B)\times C \rightarrow A\times (B\times C)$.
\end{itemize}
\end{defi}
Moggi \cite{moggi:1989} shows that the pure $\lambda$-calculus can be interpreted inside strong monads.

In the present case the natural transformation $t$ can be defined by
$$
\begin{array}{l}
t_{A,B} \colon A\times\copar{B} \rightarrow \copar{(A\times B)}\\
t_{A,B}\ \langle a,\return{b}\rangle := \return{\langle a,b\rangle}\\
t_{A,B}\ \langle a,\step{x}\rangle := \step{t_{A,B} \langle a,x\rangle}.
\end{array}
$$
\begin{thm}
The quadruple $(\copar{\_},\returnsym,\_^*,t)$ is a strong monad.
\end{thm}

A thorough study of this monad is the subject of a forthcoming article in collaboration with Thorsten Altenkirch and Tarmo Uustalu.

\section*{Acknowledgment}
During the Dagstuhl Seminar on {\em Dependently Typed Programming\/}
in the Summer of 2004, I discussed partiality in type theory with
Thorsten Altenkirch and Tarmo Uustalu.  Their comments, then and
later, were a valuable contribution to this article.  Ana Bove read
the article very carefully.  She had many insightful comments and
suggestions for improvement.  She also signalled several typos.
Thanks to her I was able to improve it considerably.  I am also
indebted to two referees who gave constructive criticism and
suggestions for improvement. 

\bibliography{biblio}
\end{document}